\pdfoutput=1
\documentclass[aps,prd,amsmath,amssymb,reprint,nofootinbib]{revtex4-2}

\usepackage[utf8]{inputenc}
\usepackage{amsmath}
\usepackage{color}
\usepackage[breaklinks]{hyperref}
\usepackage{color}
\usepackage{subfigure}
\usepackage{verbatim}
\usepackage{subfigure}
\usepackage{graphicx}
\usepackage[normalem]{ulem}

\begin{document}
\title{Stability of the non-perturbative $O(D,D)$ de-Sitter spacetime. \\
The isotropic case.
}
\author{Przemysław Bieniek, Jan Chojnacki, Jan H. Kwapisz, Krzysztof A. Meissner}
\affiliation{Faculty of Physics, University of Warsaw,\\
Pasteura 5, Warsaw, Poland}

\begin{abstract}
The existence of de-Sitter solutions and their stability within string theory became, in recent years, one of the central research questions within the string theory community. The so-called de-Sitter conjecture states that the de-Sitter vacua are unstable within string theory. \\
Here, we study the de-Sitter solutions in string theory and their stability in the time direction using the $O(d,d)$ symmetry approach allowing for the calculation of the generic form of the string theory effective action. We study the undiscussed case of the constant generalized dilaton as well as the general form of dilaton couplings. Our findings suggest that de-Sitter solutions are unstable for the constant generalized dilaton both with and without matter. Furthermore, we found that in order for the theory to be weakly coupled the general form of dilaton coupling is in the form of a cosmological constant.

\end{abstract}


\maketitle

\section{Introduction}
\subsection{The T-duality and the $O(D, D)$ symmetry.}

In theoretical physics, ``duality'' has multiple meanings. In the context of string theory, the original meaning of duality was the symmetry between s and t channels in the strong interaction S-matrix in the Veneziano amplitude \cite{Veneziano:1968yb} in the so-called dual resonance model. A spacetime duality, known as T-duality is a more recent notion, see eg. \cite{Alvarez_1995}. Speaking colloquially, it relates physical properties at large distances and short distances. It is frequently used to show an equivalence between theories with different geometries or even topologies. In the case of compactification on circle of radius $R$, the T-duality transformation is given by:
\begin{align}
R\rightarrow{}\frac{\alpha'}{R},\quad \phi\rightarrow{}\phi -\log\left(\frac{R}{\sqrt{\alpha'}}\right),
\end{align}
where $R$ is the radius of the compactified dimension, $\phi$ is the dilaton, and $\alpha'$ is an inverse of the string's tension. If $d$ dimensions are compactified on the tori of radius $R$ the T-duality is given by $O(D,D,\mathbb{Z})$, Where the $D$ is the number of compatified dimensions. For compactifications on time-dependent backgrounds, the $D = d-1$, where $d$ is the number of dimensions in the theory.

Furthermore, as found by K.A. Meissner and G. Veneziano \cite{Meissner:1991zj,MEISSNER_1991}, in $D+1$ dimensions the low-energy stringy effective action once reduced to time-dependent backgrounds possess continuous $O(D,D,\mathbb{R})$ symmetry, from now on denoted  simply as $O(D,D)$, see also \cite{Sen:1991zi, Maharana:1992my} \footnote{Note that $d$ doesn't have to be equal to $10$ or $26$ for the argument to be valid. In particular, one of the main motivations of this article is to study the behaviour of de-Sitter solutions in the non-critical string theories.}. This symmetry is also known as the generalized T-duality. The symmetry emerges in the gravitational sector of (super) string theories after including all of the massless fields appearing in the bosonic strings: scalar dilaton $\phi(x)$, metric tensor interpreted as the spin-2 graviton $G_{\mu\nu}$ and anticommuting Kalb-Ramond 2-form field $B_{\mu\nu}$. The low-energy action takes the form
\begin{align}
\label{EffectiveAction}
    I&=\frac{1}{2\kappa^2}\int d^{1+D} x\, \sqrt{-G} e^{-2\phi}(R+4G^{\mu\nu}\partial_\mu \phi \partial_\nu \phi\nonumber\\
    &-\frac{1}{12}H_{\mu\nu\rho} H^{\mu\nu\rho} ),
\end{align}
where $D$ is the number of spatial dimensions, $R$ is the Ricci scalar curvature of the metric $G_{\mu\nu}$ and $H_{\mu\nu\rho}$ is a field strength of $B_{\mu\nu}$, defined as $H_{\mu\nu\rho}:=\partial_\mu B_{\nu\rho}+\partial_\nu B_{\rho\mu}+\partial_{\rho} B_{\mu\nu}$.
For the time-dependent backgrounds, the fields ansatz is: $G_{\mu\nu} =: g_{\mu\nu}(t)$, $B_{\mu\nu}=:b_{\mu\nu}(t)$, and $\phi=:\phi(t)$.
The effective action (\ref{EffectiveAction}) may be expressed in the manifestly $O(D,D)$ invariant form:
\begin{align}
\label{invariant action}
    I=\int dt\, e^{-\Phi} \left[-\dot{\Phi}^2-\frac{1}{8} \textrm{Tr}\left(\mathcal{\dot{S}}^2\right) \right],
\end{align}
where we omit the spatial integration and put constant in front to one. The $\Phi=2\phi-\ln\sqrt{\det g} $ is the generalized (shifted) dilaton field and $g$ is the spatial metric tensor. Following the formalism of \cite{Meissner:1991zj}, $2d$-dimensional matrix $\mathcal{S}$ is defined:
\begin{equation}\label{eq:S definition}
\mathcal{S} = \begin{pmatrix}
			bg^{-1} & g - bg^{-1}b\\
				g^{-1} & -g^{-1}b
			\end{pmatrix}.\\  
\end{equation}
The global $O(D,D)$ group acts as
\begin{align}
\label{eq:Odd transformation}
    \Phi \xrightarrow{}\Phi,\quad \mathcal{S}\xrightarrow{}\Omega ^T \mathcal{S}\Omega.
\end{align}
As shown by \cite{Meissner1997} also for the $\alpha'$ corrected actions the fields can be transformed in such a way that $O(D,D)$ symmetry is manifest. Using this reasoning, in  \cite{Hohm:2019jgu}, the full, non-perturbative form of the $\alpha'$ corrected effective action on cosmological backgrounds has been derived 
\begin{align}\label{eq:hohmzwiebachaction}
		I = \int\,dt\,e^{-\Phi} \left[-\dot{\Phi}^2 + \sum_{k=1}^{\infty} \alpha'^{k-1}\left(c_k \mathrm{Tr}\left(\dot{\mathcal{S}}^{2k}\right) \right.\right.\\
		+ \textrm{multitrace}\Bigl) \Biggr].
	\end{align}
A generic multitrace term at order $\alpha'^{k-1}$ is of the following form:
\begin{equation}\label{eq:multitrace}
    \prod_i\mathrm{Tr}\left(\dot{\mathcal{S}}^{2n_i}\right)
\end{equation} 
with $\sum_in_i = k$ and $n_i > 1$. 
\subsection{The FLRW metric}
This appropach for the cosmological background \cite{Hohm:2019jgu} based on the earlier $O(D,D)$ symmetry considerations \cite{Meissner:1991zj, Meissner:1991ge, Gasperini:1991ak, Sen:1991zi, Meissner1997} opened a new window for the non-perturbative investigations of the string cosmology. 
In particular,  for the FLRW metric $g_{ij} = a^2(t)\delta_{ij}$ and vanishing Kalb-Ramond field $b_{ij}(t)=0$ the $\mathcal{S}$ reduces to
\begin{align}
    \mathcal{S} = \begin{pmatrix}
			  		0 & g \\
			  		g^{-1} & 0
			  	  \end{pmatrix}
		  	  	= \begin{pmatrix}
		  	  		0 & \mathbf{a}^{2}\\
		  	  		\mathbf{a}^{-2} & 0
		  	  	\end{pmatrix}.
\end{align}		  	  	
And its derivative is:
\begin{equation}
	\dot{\mathcal{S}} = \begin{pmatrix}
							0 & 2\mathbf{a}\dot{\mathbf{a}}\\
							-2\mathbf{a}^{-3}\dot{\mathbf{a}} & 0
						\end{pmatrix},
\end{equation}
with $\mathbf{a} = \textrm{Diag}\left(a(t), a(t), \ldots, a(t)\right)$. For this ansatz, the equations of motions derived from the action \eqref{eq:hohmzwiebachaction} are given by
\begin{align}
\label{eq:non-perturbative EOM}
2 \ddot{\Phi} - \dot{\Phi}^2 +F(H)=0, \nonumber\\
f'(H) \dot{H} - \dot{\Phi} f(H) =0,\nonumber\\
\dot\Phi^2+g(H)=0.
\end{align}
where the $'$ denotes the derivative with respect to the Hubble parameter $H=\dot{a}/a$, where
\begin{align}
    F(H) = 2 d \sum_{k=1}^{\infty} (-\alpha')^{k-1} c_k 2^{2k} H^{2k}.
\end{align}
and 
\begin{align}
  f(H) := F'(H),\quad g(H) := HF'(H)-F(H).
\end{align}
\subsection{de-Sitter solutions}
The existence of de-Sitter solutions and their stability within string theory became, in recent years, one of the central research questions within the string theory community. The so-called de-Sitter conjecture states that the de-Sitter vacua are unstable within string theory \cite{Obied:2018sgi} \footnote{The original de-Sitter conjecture stated that there are no de-Sitter vacua \cite{Obied:2018sgi}, while the so-called refined de-Sitter conjecture states the instability of the de-Sitter vacua \cite{Garg:2018reu, Ooguri:2018wrx}. In the following text, we will use the term de-Sitter conjecture in the sense of the refined de-Sitter conjecture. Overall there is no no-go theorem, however the ``acceptable`` de-Sitter solutions are hard to be found.}.\\
The general form of the string theory effective action and the associated equation \eqref{eq:non-perturbative EOM} revealed a new possibility of a non-perturbative de-Sitter solution once summing all the possible $\alpha'$ corrections in the $O(D, D)$ formalism \cite{Hohm:2019jgu}. The existence of such solutions has been also investigated in the presence of matter \cite{2020_Brandenberger}, see also for the study in the Einstein frame \cite{Krishnan:2019mkv}. Furthermore based on the $\alpha'$-complete equations found by Hohm and Zwiebach, it was shown in \cite{Wang:2019kez} that there exists a solution smoothly connecting pre-Big Bang and post-Big Bang cosmology.\\ 
Assuming that the form of the effective action takes a suitable form \footnote{Let us note that the $c_k$ coefficients are known only up to the very low order, in particular $c_1=- \frac{1}{8}$, such that $F(H) = - d H^2 $ at the zeroth order in $\alpha'$. In particular, in \cite{Hohm:2019ccp} the authors postulated that the $c_i$ can be such that the de-Sitter solution can be found. This claim has been investigated with the functional renormalization group calculations \cite{Basile:2021euh, Basile:2021krk, Basile:2021krr} yielding a negative answer and thus no de-Sitter solutions have been found when resorting to the truncations of the effective action or in low dimensions. It argued, that in the heterotic theory it is not possible to construct a de-Sitter vacuum at tree-level in $g_s$ \cite{Kutasov_2015}, therefore in this theory the values of the constants are most likely unsuitable. Below we will assume that the coefficients $c_k$ are s.t. the de-Sitter solutions exist.} such that a de-Sitter solution can be obtained and that the resulting cosmological constant can be comparable to the observational one, the next step is to study the stability of the solutions.  The studies  \cite{2020_Bernardo, 2021Nunez} revealed that when the generalized dilaton decays the de-Sitter solutions are generally stable. However, authors of \cite{2020_Bernardo} were studying the stability in presence of matter, while the authors of \cite{2021Nunez} briefly mentioned that a separate case for constant generalized dilaton needs to be considered, see Sec~3.4 in \cite{2021Nunez} for further details.
 \\
 \paragraph*{Summary of work}
In this work, we extend the previous results in several directions. In Sec.~\ref{sec:de-Sitterstability} we study the case of the $\dot{\Phi}(H_0)=0$ in the absence of matter that was not considered in \cite{2021Nunez} and show that the de-Sitter solutions are unstable for that case. Constant generalized dilaton on de Sitter spacetimes leads to the "rolling dilaton" scenario often encountered in the string cosmology \cite{BERNDSEN_2004,MUELLER199037} and furthermore for de-Sitter solution they correspond to the linear dilaton CFTs, being the solutions in the non-critical strings \cite{POLYAKOV1981207}. Time dependent dilaton is suspected as a mechanism governing the radii of the compactified dimensions. Recently, the rolling dilaton has been found necessary for connecting epochs in the string gas cosmology models \cite{Bernardo_2021}.

We present our findings on the exemplary solution discussed in \cite{Hohm:2019jgu, Hohm:2019ccp}. In \ref{sec:dilatonpotential} we extend the previous studies of inclusion of matter coupled in the $O(d,d)$ fashion. In particular, we study the general dilaton potential. Our results imply that the generic form of dilaton potential is of the cosmological constant case in \ref{sec:cosmoconstant} if the theory is to be weakly coupled (in the sense of the dilaton coupling). This is important in the late-time evolution of the universe, dominated by the cosmological constant. Cosmological scenario with a constant generalized dilaton would not lead to an indefinitely inflating universe, in contrast with the current observations.\\
Here we study the stability of the FLRW solutions in the time direction. Furthermore, in \cite{Paper2} we are studying the anisotropic dynamics perturbatively at orders up to $\alpha'^3$ and their dependence on the string theory type. Our findings suggest, that at those orders no stable isotropic de-Sitter solution is possible.
\section{Stability analysis of the vacuum de-Sitter solutions for $\dot{\Phi} = 0$.}

\label{sec:de-Sitterstability}
As discussed in the introduction, we shall study the case of constant generalized dilaton at the de-Sitter solution with Hubble constant $H_0$, \textit{i.e.} $\dot\Phi|_{H=H_0}=0$, not discussed in \cite{2021Nunez}. Before we start our analysis, let us note that for the FLRW metric the ordinary dilaton, satisfies the equation below:
\begin{align}
    \Phi =2\phi - \log a(t)^3, 
\end{align}
and $\Phi := \Phi_0$. For the de-Sitter solution $a(t) = e^{H_0 t}$, we get:
\begin{align}
    \Phi_0 =2 \phi - 3 H_0 t, 
\end{align}
and $\phi = \frac{1}{2}\Phi_0 + \frac{3}{2} H t$, which is a well known linear dilaton solution in non-critical string theories.
We start our analysis by rewriting the equations of motion \eqref{eq:non-perturbative EOM} in a way convenient for the stability analysis. The third equation of \eqref{eq:non-perturbative EOM} implies:
\begin{align}
\label{eq:dotphi}
    \dot\Phi=\pm \sqrt{-g(H)},
\end{align}
where we assume  that $-g(H)$ is a non-negative function. The first equation of \eqref{eq:non-perturbative EOM} implies 
\begin{align}
-\dot{\Phi} f(H) + f'(H) \dot{H}=0
\end{align}
We shall study two separate cases.
\subsection{ Case 1: $f'(H)\neq 0$ around $H=H_0$.}
\begin{figure}[t!]
\includegraphics[width=0.49\textwidth]{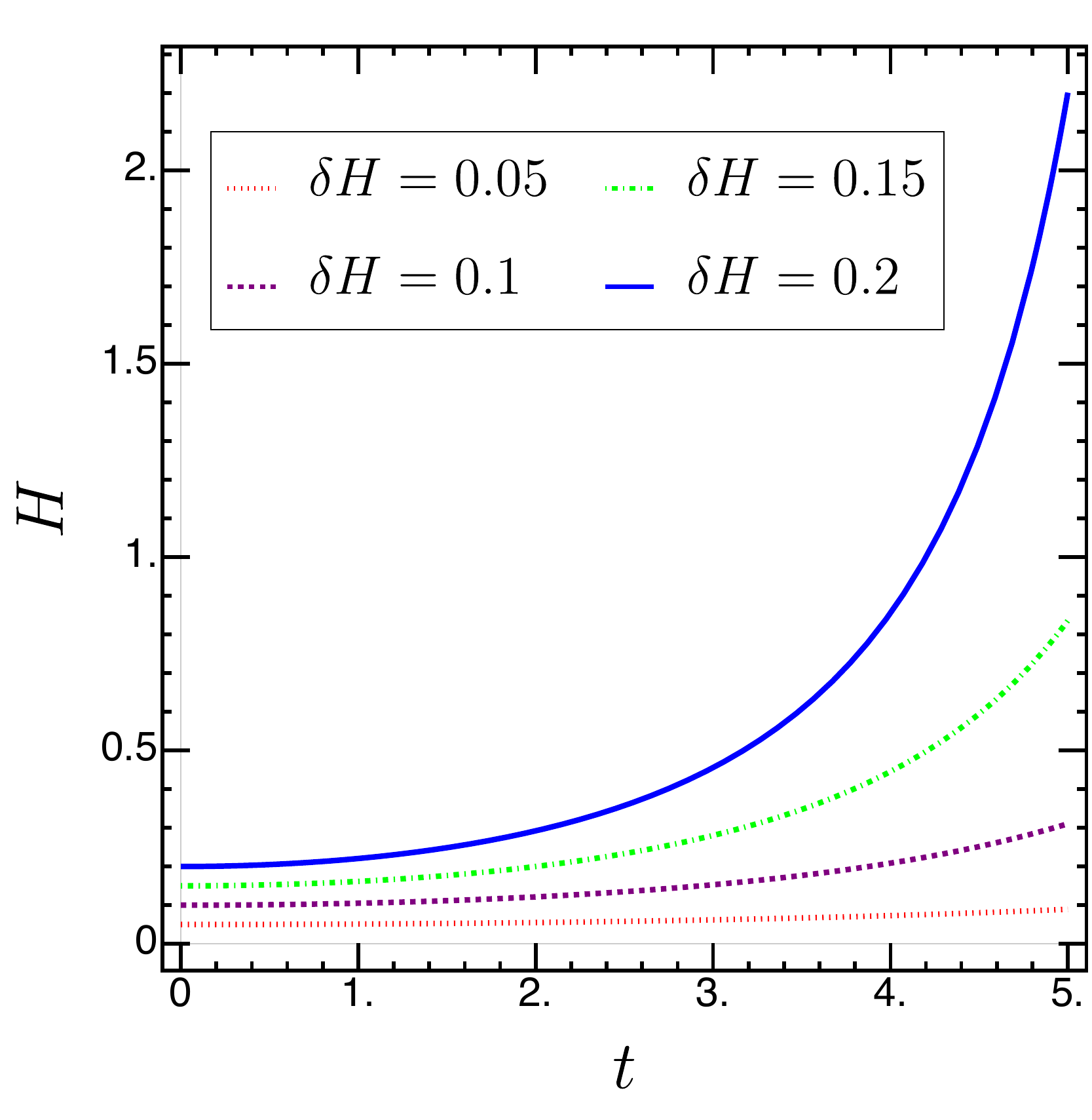}
\label{fig:evolution}
\caption{Examples of the unstable evolution with initial conditions $\dot \delta H(0)=0$ 
and $\delta H=\{0,0.05,0.1,0.15,0.2\}$.}
\end{figure}

Using $f'(H)\neq 0$ together with \eqref{eq:dotphi} we find an expression for $\dot H$:
\begin{align}
\label{eq:dotH}
    \dot H=\pm \frac{\sqrt{-g(H)}f(H)}{f'(H)},
\end{align}
Differentiating equation \eqref{eq:dotH} and substituting back $\dot H$ gives us:
\begin{align}
\label{eq:ddotH}
    \ddot H=-\frac{1}{2}\frac{H f(H)^2}{f'(H)}-\frac{g(H)f(H)}{f'(H)}+\frac{g(H)f''(H)f(H)^2}{f'(H)^3}.
\end{align}
As discussed in the introduction, we shall study the case of constant generalized dilaton at the de-Sitter solution, \textit{i.e.} $\dot\Phi|_{H=H_0}=0$, not discussed in \cite{2021Nunez}. This by (\ref{eq:non-perturbative EOM}, \ref{eq:dotH}) leads to a condition $f(H_0)=g(H_0)=0$. Furthermore, notice that $\ddot{H}|_{H=H_0}=0$. In order to study the stability of Eq.~\eqref{eq:ddotH} we perform the power series expansion around $H_0$ of the rhs of \eqref{eq:ddotH}\footnote{A curious reader might wonder why the \eqref{eq:ddotH approx} is no-longer the $O(D,D)$ symmetric. It is so due to the fact that we are expanding around a non-trivial vacuum $H_0$, similarly as in the Higgs mechanism, where expansion around non-trivial minima triggers the spontaneous symmetry breaking. Furthermore it is natural since originally the $O(D, D)$ symmetry was discovered as the solution generating techniques \cite{Meissner:1991ge,Meissner:1991zj}. We are planning to investigate that in the future in more depth.}:
\begin{align}
\label{eq:F}
    \ddot H \equiv W(H)= W(H_0)+W'(H_0)\left(H-H_0\right)\\
    +\frac{1}{2!}W''(H_0)\left(H-H_0\right)^2+\dots
\end{align}
The first derivative of $W(H)$ is:
\begin{align}
\label{eq:F'}
    W'(H)&=-2 H f(H)-g(H)-\frac{3 f(H)^2 g(H) f''(H)^2}{f'(H)^4}+\\
    &+\frac{3 f(H) g(H) f''(H)}{f'(H)^2}+\frac{3 H f(H)^2 f''(H)}{2
   f'(H)^2}-\frac{f(H)^2}{2 f'(H)}\nonumber\\&+\frac{f^{(3)}(H) f(H)^2 g(H)}{f'(H)^3}.\nonumber
\end{align}
Keeping in mind that $f(H_0)=g(H_0)=0$, we can see that the $f'(H_0)=0$ and hence the first order of the expansion vanishes. Furthermore, the only non-vanishing contribution to $W''(H_0)$ comes from the first two terms in \eqref{eq:F'}. All of the remaining terms will be proportional to either $f(H)$ or $g(H)$:
\begin{align}
    W''(H_0)=\frac{d}{d H}\Big|_{H=H_0} \left(-2H f(H)-g(H) \right)+vanishing.
\end{align}
Finally, using $g'(H)=Hf'(H)$ we find:
\begin{align}
\label{eq:ddotH approx}
    \ddot H =-\frac{3}{2} H_0 f'(H_0)\left(H-H_0\right)^2+\mathcal{O}(H-H_0)^3.
\end{align}
Having obtained the first non-vanishing contribution we can start to study the stability of the de-Sitter solutions with $\dot{\Phi}(H_0)=0$.
If we denote $\delta H = H-H_0$ and $a=-\frac{3}{2} H_0 f'(H_0)$ we obtain the following equation
\begin{align}
\label{eq:EOM at H0}
\ddot {\delta H}&=a (\delta H)^2,
\end{align}
which resembles a particle in a cubic potential
\begin{align}
\label{eq:EOM_EP}
\ddot {\delta H}&=-\nabla V(\delta H),
\end{align}
with $V(\delta H) = \frac{a}{3} (\delta H)^3$ and the field initially settled in $H=H_0+\delta H$ with $\delta H \ll 1$ will inevitably ``roll down" the slope. This is independent of the value and sign of $a$, given that $H_0$ and $f'(H_0)$ are non-zero. Thus de-Sitter vacuum is unstable, because the first-order perturbation $F'(H_0)$ is vanishing.
We may solve equation \eqref{eq:EOM at H0} numerically for the initial conditions $\dot{\delta H}(0)=0,\quad \delta{H}(0)=\epsilon$. This is depicted on Fig.  \ref{fig:evolution}.\\ 
\begin{figure}[t!]
\includegraphics[width=0.46\textwidth]{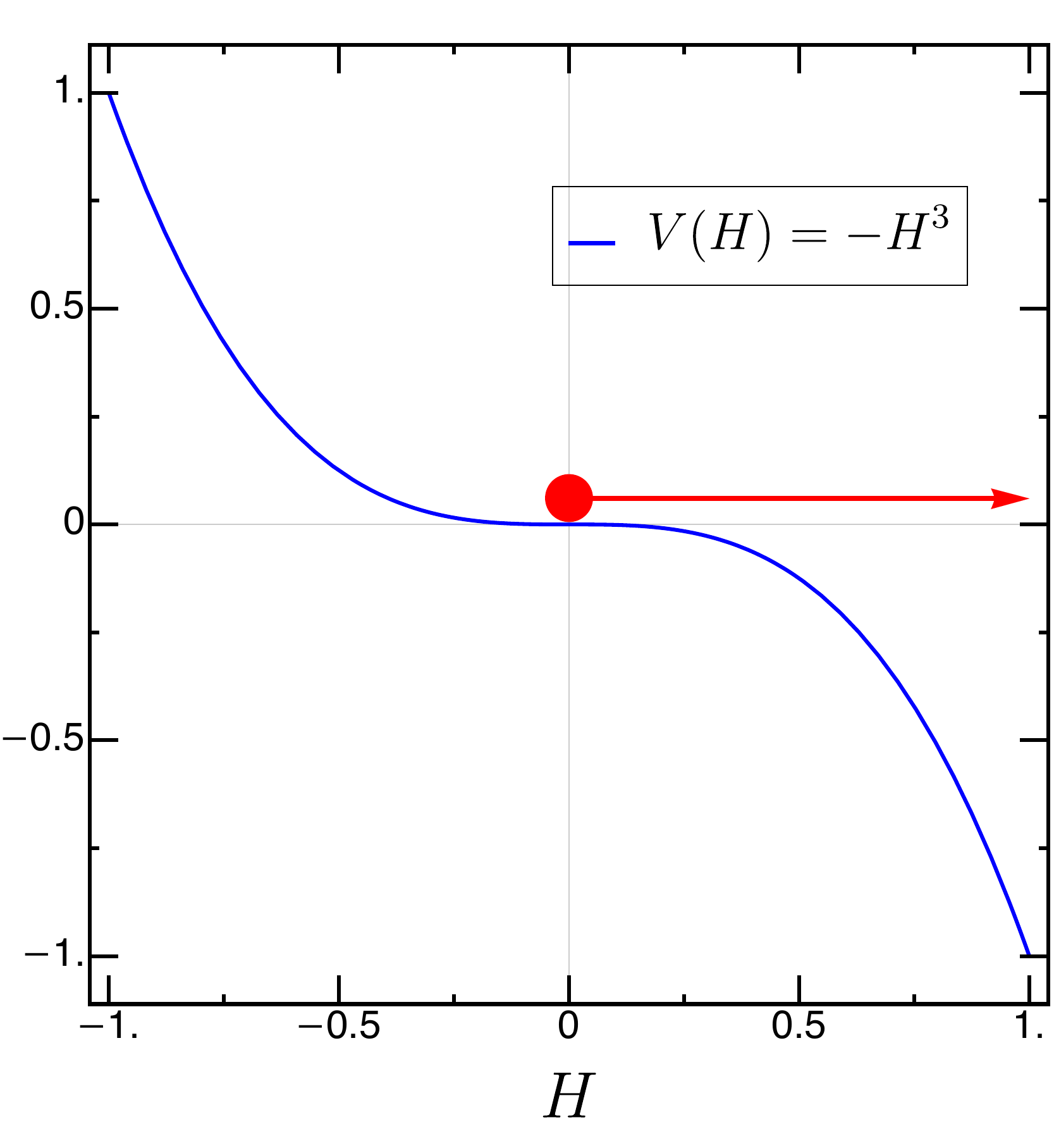}
\label{fig:potential}
\caption{The cubic potential around the de-Sitter point $H=H_0$}
\end{figure}
\subsection{Case 2: $f'(H)|_{H=H_0} =0$.}
In the case of $f'(H_0)=0$, we cannot apply the previous reasoning directly, since the denominator in \eqref{eq:dotH} vanishes. However, we will still be able to rewrite the EOM in the effective-potential form \eqref{eq:EOM_EP}. To show this, we start with the equation \eqref{eq:dotH} multiplied by its denominator:
\begin{align}
\label{eq:dotH2}
    \mp \sqrt{-g(H)}f(H)+f'(H)\dot H=0.
\end{align}
Now, however, not only $f(H_0)$ and $g(H_0)$ vanish, but also their first derivatives $f'(H_0)=g'(H_0)/H_0=0$. We are left with the Taylor expansions of $f$ and $g$ around $H_0$ starting at the second order:
\begin{align}
    g(H)&=\frac{1}{2}g''(H_0)\left(H-H_0\right)^2+\dots\\
    f(H)&=\frac{1}{2}f''(H_0)\left(H-H_0\right)^2+\dots.\nonumber
\end{align}
We may insert these in \eqref{eq:dotH2}, to get:
\begin{align}
    \dot{\delta H} =a \delta H^2,
\end{align}
where, as before, $\delta H=H-H_0$ and $a=\pm\frac{1}{2}\sqrt{-\frac{1}{2}g''(H_0)}$. Thus the evolution around the de-Sitter point is unstable, regardless of the sign of $a$. 
\subsection{Exemplary $f(H)$ function}
\label{sec:examples}
\begin{figure}[h!]
\includegraphics[width=0.497\textwidth]{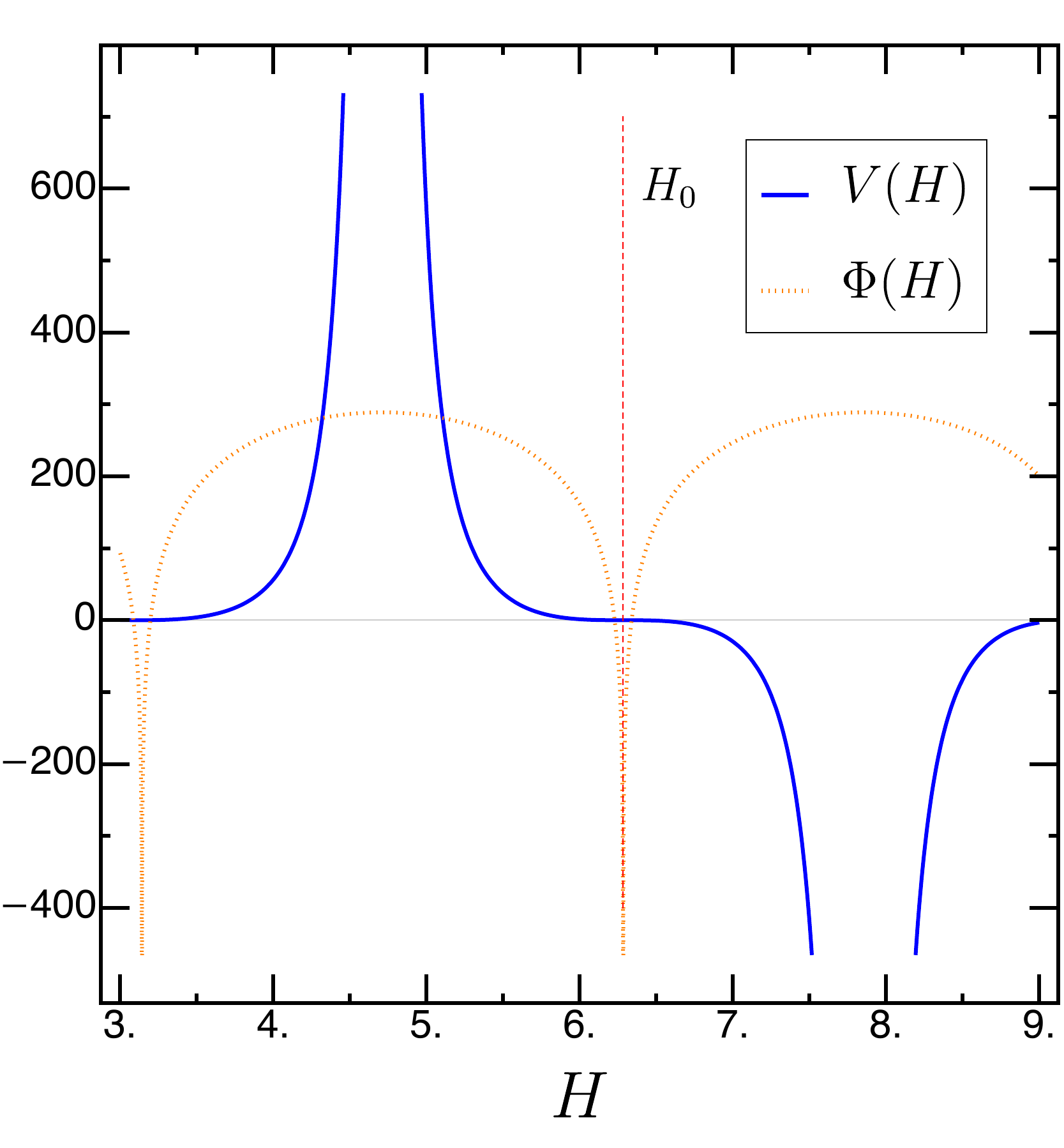}
\caption{Schematical evolution around de-Sitter point in the sine solution. The blue curve corresponds to the potential, while the yellow one gives the dilaton value for a particular $H$. Hubble parameter will ``move" from the region of higher potential towards the region of lower potential. Starting at the unstable point $H_0$, the evolution rolls towards the ``well". Around the de-Sitter point, the value of the dilaton changes rapidly.}
\label{fig:potential example}
\end{figure}

As a case study of the de-Sitter stability we consider the sine solution found in \cite{Hohm:2019jgu}:
\begin{align}
\label{deSitter example}
    f\left( H\right)&=-\frac{2d}{\sqrt{\alpha'}}\sin\left(\sqrt{\alpha'}H\right),\\
    g\left(H\right)&=-\frac{2d}{\sqrt{\alpha'}}\left(\sqrt{\alpha'}H\sin\left(\sqrt{\alpha'}H\right)+\cos\left(\sqrt{\alpha'}H\right)-1\right).\nonumber
\end{align}
We would like to see, whether the approximations leading to \eqref{eq:ddotH approx} are justified. Since \eqref{eq:ddotH} is an exact result, we may insert the above de-Sitter solution to \eqref{deSitter example}:
\begin{align}
\label{ddotH example}
   \sqrt{\alpha'} \ddot { H}=-\frac{d}{8 \alpha'} \sec ^3\sqrt{\alpha'} H\left(-9 \sqrt{\alpha'} H+16 \sin \sqrt{\alpha'} H\right.\nonumber\\
   -8 \sin \left(2 \sqrt{\alpha'} H\right)+8 \sqrt{\alpha'} H  \cos \left(2 \sqrt{\alpha'} H\right)\nonumber\\
   \left.+8 \sqrt{\alpha'} H\cos \left(4 \sqrt{\alpha'} H\right)\right).
\end{align}
 Discrete infinity of de-Sitter solutions found in \cite{Hohm:2019jgu} is $\sqrt{\alpha'} H_0= 2\pi n$. Power-series expansion of $\ddot { H}$ around $ H_0=\frac{2\pi}{ \sqrt{\alpha'}}$ begins with the second-order term, as expected:
\begin{align}
\label{ddotH example approx}
    \ddot { H}&=6 \pi  d \left(\sqrt{\alpha'} H-2 \pi \right)^2 +\mathcal O\left(\sqrt{\alpha'} H-2 \pi \right)^3,
\end{align}
the potential defined in \eqref{eq:EOM at H0} is cubic and the solution is unstable. This example sheds more light on the  generalized dilaton behaviour around the de-Sitter point. Using the first equation in \eqref{eq:Dilaton}, we find it as a function of $\bar H$:
\begin{align}
    \Phi \left( \sqrt{\alpha'} H\right)=\alpha'\ln\left|\sin  H \right|+const,
\end{align}
where the constant stems from the initial conditions. Since integrating out the exact equation \eqref{ddotH example} or the approximate equation \eqref{ddotH example approx} is neither easy nor instructive, we will restrain ourselves to a qualitative analysis. The initial value of the Hubble parameter is set around the de-Sitter point $H_0$. The effective potential picture suggests, that the field will roll down from this point, as it may be seen on Figure \ref{fig:potential example}. 
The cubic behaviour of the potential around $ H_0=\frac{2\pi}{\sqrt{\alpha'}}$ is clear. At $ H=\frac{5}{2}\frac{\pi}{\sqrt{\alpha'}}$ dilaton has a plateau, and the potential is singular. We may wonder, whether it is another de-Sitter point. This is not the case, since the last equation in \eqref{eq:non-perturbative EOM}
yields non-zero $f(\frac{5}{2\sqrt{\alpha'}}\pi)\neq 0$. Moreover, $\dot\Phi\neq0$ yielding our previous approximate analysis inapplicable.\\
In general, neither $f(H)$, nor the potential needs to be periodic. We claim, however, that the evolution of the dilaton around $H_0$ will be similar. This follows from a simple fact, that the equation \eqref{eq:Dilaton} is singular for $f(H)=0$, which is of course true at $H=H_0$. Hence, the dilaton will change rapidly, when departing from $H_0$ slightly.
\section{de-Sitter solution with dilaton potential}
\label{sec:dilatonpotential}
\subsection{General case}
Here, we extend the previous study by including the matter. In string theory, the low-energy effective action consist of $\phi, b_{\mu\nu}$ and $g_{\mu\nu}$ on the fundamental level. Those matter fields might be some remnants of the compactification of the latter fields from $D$ to $d$ dimensions or string gas stemming from the massive modes of the strings, modelled as the string gas cosmology \cite{BRANDENBERGER1989391, TSEYTLIN1992443, Gasperini:1991ak}. Otherwise, one might just introduce the matter sources phenomenologically without specifying their origin.
Recently, a general way of $O(d, d)$ invariant matter coupling was developed in \cite{2020_Brandenberger}.\\
Given the matter action $S_m \left[\Phi, n, \mathcal{S}, \chi \right]$ of the matter field(s) $\chi$, the $\alpha'$-complete Friedmann equations acquire source terms:
\begin{align}
\label{eq:matter sources}
    -\sqrt{-g}\rho=\frac{\delta S_m}{\delta n},\quad \sigma=-\frac{2}{\sqrt{-G}}\frac{\delta S_m}{\delta \Phi},
\end{align}
where $\rho$ is the energy density and $\kappa$ is a coupling constant.
There is a similar equation for the relativistic pressure $p$, it is however more convenient to calculate it from the continuity equation:
\begin{align}
\label{eq:continuity}
    \dot {\tilde {\rho}}+dH\tilde p-\frac{1}{2}\dot\Phi\tilde\sigma=0,
\end{align}
where
\begin{align}
    \tilde\rho=e^{\Phi}\sqrt{g}\rho,\quad \tilde\sigma=\frac{1}{2}e^{\Phi}\sqrt{g}\sigma,\quad \tilde p= \frac{1}{d}e^{\Phi}\sqrt{g}p.
\end{align}
 The $\alpha'-$complete equations of motion, coupled to matter were found in \cite{2020_Brandenberger}:
\begin{align}
\label{eq:EOM_matter}
    \dot\Phi^2+g(H)&=2\kappa^2 \tilde\rho,\\
    f'(H)\dot H-\dot\Phi f(H)&=-2d\kappa^2  \tilde p,\nonumber\\
    2\ddot\Phi -\dot \Phi^2+F(H)&= 2\kappa^2  \tilde \sigma.\nonumber
\end{align}
Let us first note that since $\tilde{\rho}<0$ (see below), $\dot{\Phi}^2>0$ one requires $g(H)$ to be negative. One possible string inspired matter source could be the dilaton potential $V(\Phi)$ (all the kinetic terms can be combined into the $O(d, d)$ action \cite{2020_Brandenberger}). Using the (\ref{eq:matter sources}, \ref{eq:continuity}) one obtains:
\begin{align}
  \tilde{\rho} = -V(\Phi), \quad \tilde{p} =0,\quad \tilde \sigma = V(\Phi) - \frac{\partial V}{\partial \Phi}.
\end{align}
Given de-Sitter solution $H=H_0$ one obtains:
\begin{align}
\label{eq:deSitterpotential}
    \dot\Phi^2+g(H_0)& = -2 V(\Phi),\\
    -\dot\Phi f(H_0)&= 0,\nonumber\\
    2\ddot\Phi -\dot \Phi^2+f(H_0)H - g(H_0)&= 2\left(V(\Phi)- \frac{\partial V}{\partial \Phi}\right),\nonumber
\end{align}
where we put $\kappa=1$. 
From the second equation of \eqref{eq:deSitterpotential} we have two separate cases $f(H_0) =0$ or $\dot{\Phi}=0.$ For $\dot{\Phi}=0$ the equations are contradictory unless $\frac{\partial V}{\partial\Phi}=0$ unless the constant potential.\\
For $f(H_0)=0$ we get 
\begin{align}
\label{eq:fH0dilaton}
    \dot\Phi^2+g(H_0)& = -2 V(\Phi),\\
    2\ddot\Phi -\dot \Phi^2- g(H_0)&= 2\left(V(\Phi)- \frac{\partial V}{\partial \Phi}\right),\nonumber
\end{align}
adding the equations we get the equation for the dilaton evolution as
\begin{align}
\label{eq:dilatonevolution}
    \ddot\Phi =- \frac{\partial V}{\partial \Phi}.
\end{align}
together with the constraint equation for $\dot{\Phi}$
\begin{align}
\label{eq:dilatonconstraint}
     \dot\Phi^2& = -2 V(\Phi)-g(H_0).  
\end{align}
Note that for the above equation to hold, one requires $-1/2g(H_0) > V(\Phi)$,
thus the potential has to be to bounded from above. Secondly note that the \eqref{eq:dilatonevolution} is the equation for the particle in potential $V(\Phi)$.  We have inspected the numerical solutions of the \eqref{eq:dilatonevolution} for potential of the form $V(\Phi)= - c_n \Phi^{2n}$, for $n\in (2,\ldots,10)$ as well as $V(\Phi) = - c \exp (\Phi)$ with $c>0, c_n>0$. We found that the generalized dilaton necessary develops a negative value during the time evolution (the special case of $n=1$ will be discussed below) and furthermore at some point $\Phi < -d H t$. On the other hand, in order for the theory to be weakly coupled one requires $e^{-\phi}\ll1 $ and thus $\phi>0$, where $\phi = \Phi + d H t$ for the de-Sitter solutions. In particular, for Minkowski spacetime $\phi = \Phi$. Thus for the de-Sitter spacetime the theory becomes strongly coupled for $n>1$. \\
Let us now discuss the special case of  $V(\Phi) = -\frac{1}{2} m^2 \Phi^2$ the analytical solution of \eqref{eq:dilatonevolution} is $\Phi = \mathbf{c}_1 \cosh(m t) + \mathbf{c}_2 \sinh(m t).$ Plugging this solution into the \eqref{eq:dilatonconstraint} one gets $m=0$.\\
We conclude that for the weakly coupled theory, in order to have de-Sitter solution, one requires  $ \frac{\partial V}{\partial \Phi}=0.$ We shall now discuss the cosmological constant case, where $V(\Phi) = \Lambda$.

\subsection{The cosmological constant case}
\label{sec:cosmoconstant}
The cosmological constant term given by $\Lambda = \frac{d-d_{\mathrm{crit}}}{\alpha'}$ is in fact required for the noncritical strings to remain Weyl invariant and together with $\dot{\Phi} = 0$ is natural extension of the formalism discussed above to the non-critical strings case.
For the cosmological constant case from equations \eqref{eq:matter sources} and \eqref{eq:continuity} we find:
\begin{align}
    \tilde\rho=-\Lambda,\quad \tilde\sigma=2\Lambda,\quad \tilde p=0.
\end{align}
Leading to equations of motion:
\begin{align}
    \dot\Phi^2+g(H)&=-2 \Lambda,\\
    f'(H)\dot H-\dot\Phi f(H)&=0,\nonumber\\
    2\ddot\Phi -\dot \Phi^2+H f(H)-g(H)&=2 \Lambda.\nonumber
\end{align}
de-Sitter solution is now given by $g(H_0)=-2 \Lambda$ and as before we have either $\dot{\Phi}=0$ or $f(H_0)=0$. In the $\dot{\Phi}=0$ case, we get
\begin{align}
    g(H_0)&=-2 \Lambda,\\
  H f(H_0)-g(H_0)&=2 \Lambda.\nonumber
\end{align}
and thus $f(H_0)=0$. Let us proceed the stability of the $f(H_0)=0$ case. Similarly to \eqref{eq:dotH}, we find the first derivative of Hubble parameter
\begin{align}
    \dot H=\pm \frac{f(H)}{f'(H)}\sqrt{-2\kappa^2 \Lambda-g(H)}.
\end{align}
It is vanishing at $H=H_0$. The rest of the reasoning leading to \eqref{eq:ddotH approx} remains unchanged. 
This is true, however, for a constant value of the dilaton first derivative $\dot\Phi=0$. As previously described in \cite{2020_Bernardo}, for the general, time dependent $\Phi(t)$, we have the stable branch of solutions with $\dot\Phi<0$ and the unstable branch $\dot\Phi>0$. The instability is present, regardless of the sign of the cosmological constant. \\
The generalized dilaton is also unstable around the de-Sitter solution. To investigate it, let us start the evolution around a de-Sitter solution $H_0$ by adding a small perturbation $\Delta$ to it. As we have discussed before, $H$ will grow with time. We further assume a general power-law behaviour of the Hubble parameter and approximate: 
\begin{align}
    f(H)&=A\left(H-H_0\right),\\
    H&=H_0+\Delta+at^n.
\end{align}
From the first equation in \eqref{eq:non-perturbative EOM} we have
\begin{align}
 \dot\Phi=\frac{f'(H)}{f(H)}\dot H,   
\end{align}
After integrating over time and changing the variables we are left with
\begin{align}
\label{eq:Dilaton}
    \Phi\left(t\right)=\ln\left|\frac{f(H)}{f(H_0+\Delta)}\right|=\ln\left|1+\frac{a}{\Delta}t^n\right|.
\end{align}
Clearly, the dilaton grows unboundedly with time. which means that the system of differential equations is unstable in both "directions" $H$ and $\Phi$.
We wish to explain how our work relates to the investigations of \cite{2020_Bernardo}. The authors present a thorough discussion of stability around equilibria for a general equation of state. They find eigenvalues of the matrix describing the linearized system \eqref{eq:EOM_matter}:
\begin{align}
    \dot H &= \frac{1}{F''} \left(y F' - d \omega (y^2+HF' - F) \right),\\
    \dot y &= \frac{y^2}{2}-\frac{F}{2}+\frac{\lambda}{4}(y^2+HF'-F).\nonumber
\end{align}
Here, $F'(H)=f(H)$, $y=\dot\Phi$, $\omega= p/\rho$, $\lambda=\sigma/\rho$. The linearization matrix $A=\partial_{H,y} C_i(X)$ is then
\begin{align}
  A=  \begin{pmatrix}
y_0-dH_0\omega & \frac{1}{F''_0}(F'_0-2d\omega y_0) \\
-\frac{F'_0}{2}+\frac{\lambda}{4}H_0F''_0 & y_0(1+\frac{\lambda}{2})
\end{pmatrix},
\end{align}
where the subscript zero relates to the values at the de-Sitter point $H=H_0$. In the cosmological constant case: $\omega=0,\quad \lambda=-2,\quad F'_0=f(H_0)=0$. The linearization matrix simplifies greatly:
\begin{align}
  A=  \begin{pmatrix}
y_0 & 0 \\
-\frac{1}{2}H_0F''_0 & 0
\end{pmatrix}.
\end{align}
The eigenvalues are $\alpha_{\pm}=\{0,y_0\}$, so the stability of the system depends on $y_0=\dot\Phi(H_0)$. If it was negative, the system would be stable. From the second equation in \eqref{eq:EOM_matter} evaluated in $H=H_0$ we have:
\begin{align}
    \frac{d}{dt}\left(e^{-\Phi}f(H_0)\right)=0,
\end{align}
hence $\dot\Phi(H_0)=0$. Since both of the eigenvalues are zero, the linear analysis is not sufficient to conclude that the point $H_0$ is stable. The second-order analysis described in the previous section shows that the de-Sitter vacuum is unstable. Nevertheless, the discussion of the solution with $\dot\Phi(H_0)\neq 0$ in \cite{2020_Bernardo} is still valid and leads to a stable de-Sitter vacuum for $\dot\Phi(H_0)< 0$. 
\section{Conclusions}
In this paper we have studied the stability of the de-Sitter solutions in string theory using the non-perturbative effective action derived within the $O(d,d)$ symmetry. In particular, we have studied the $\dot{\Phi}=0$ vacuum solution, resembling the non-critical strings linear dilaton CFT solution, using the non-linear analysis and an exemplary solution $f(H) \sim \sin\left(\sqrt{\alpha'} H\right)$. For the matter case we have shown that in order for the theory to be weakly coupled during evolution one requires the generic dilaton potential to be constant. We have built a simple framework employing the idea of the effective potential for the stability analysis in the T-dual theories. The approach is intuitive and less tedious, than more conventional expansion of the EOMs to the second-order variations. Our results concerning the constant generalized dilaton greatly impact de Sitter theories with rolling dilaton, yielding them unstable.

\addcontentsline{toc}{section}{The Bibliography}
\acknowledgments
J.C. thanks Robert Brandenberger and Heliudson Bernardo for numerous engaging discussions on the stability of de Sitter spacetime. J.C. is thankful for the hospitality of the McGill University.
J.H.K thanks Massachusetts Institute of Technology (MIT) for hospitality and support during this work. J.H.K. was supported by the Polish National Science Centre grant 2018/29/N/ST2/01743. K.A.M. and J.H.K. were partially supported by the Polish National Science Centre grant UMO-2020/39/B/ST2/01279
\addcontentsline{toc}{section}{The Bibliography}
\bibliography{References.bib}{}
\bibliographystyle{apsrev4-2}
\end{document}